# Geospatial-Temporal Sensemaking of Remote Sensing Activity Detections with Multimodal Large Language Model

David F. Ramirez*[ab], Tim Overman[b], Kristen Jaskie[b], Andreas Spanias[a]
[a]SenSIP Center, School of ECEE, Arizona State University, Tempe, Arizona, USA;
[b]Prime Solutions Group Inc, Goodyear, Arizona, USA

## ABSTRACT

Understanding how human activity changes the Earth over time is an important goal for modern remote sensing and geospatial intelligence. Repeated satellite observations can reveal these changes, but such observations are often sparse, irregularly sampled, affected by sensing conditions, and difficult to interpret as coherent activity sequences. Heavy construction is a useful example of this challenge because a site may progress from land clearing to site preparation, to active construction, and eventually to completion over months or years. This paper presents ongoing work toward multimodal large language model (MLLM)-based geospatial-temporal sensemaking by deriving SMART-HC-VQA, a Sentinel-2-based visual question answering (VQA) dataset from the Intelligence Advanced Research Projects Activity (IARPA) Space-Based Machine Automated Recognition Technique Heavy Construction (SMART-HC) dataset. SMART-HC-VQA converts construction-site annotations, construction-type labels, temporal phase labels, location metadata, and observation relationships into natural-language questions and answers. This formulation reframes heavy-construction monitoring as a temporally extended automatic target recognition (ATR) and VQA problem in which a fixed geospatial site is treated as a target whose attributes and activity state evolve over time. The current dataset includes 21,837 accessible Sentinel-2 image chips, 65,511 single-image VQA examples, and approximately 2.3 million two-image temporal-comparison examples generated through our novel Image-Pairwise Combinatorial Augmentation method. We describe the data preparation workflow for retrieving Sentinel-2 imagery, resolving historical image references, chipping large satellite tiles into site-centered images, preserving traceability to SMART-HC annotations, and analyzing distributions of site size, observation count, temporal coverage, construction type, and phase labels. We also describe an implemented multi-image MLLM training framework based on the LLaVA-NeXT Mistral-7B architecture, adapted to accept multiple dated image inputs and train on metadata-derived VQA examples. Rather than presenting final model evaluation results or claiming a finished operational system, this work provides a reproducible foundation for understanding language-guided remote-sensing activity, where the objective is not only to detect change but also to explain what is happening, how it is progressing, and what it may become.

## 1. INTRODUCTION

Understanding how human activity changes the Earth over time is a central challenge for modern remote sensing and geospatial intelligence. Satellites, radar systems, and other overhead sensors generate large volumes of imagery, but the analytic value of these data collections depends on more than detecting objects or identifying isolated changes. Analysts often need to determine what activity is occurring, how it is progressing, and what it may indicate about a future state, operational relevance, or human intent.

Heavy construction is a useful case study for this problem because it is spatially localized, temporally extended, and visually expressed through observable stages such as land clearing, site preparation, active building, and completion. Recent work on the Intelligence Advanced Research Projects Activity (IARPA) Space-Based Machine Automated Recognition Technique Heavy Construction (SMART-HC) dataset [1] formalizes this challenge as a broad area search and phase classification over multi-source, multi-temporal satellite image sequences. In this setting, algorithms must localize construction sites, classify activity phases, and associate observations through time. This paper builds on that formulation by examining how multimodal large language models (MLLMs) may support geospatial-temporal interpretation of construction activity, with emphasis on dataset derivation, implementation, and proposed modeling methods rather than a completed operational system.

*dframire@asu.edu; phone +1 623 853-0829; psg-inc.net

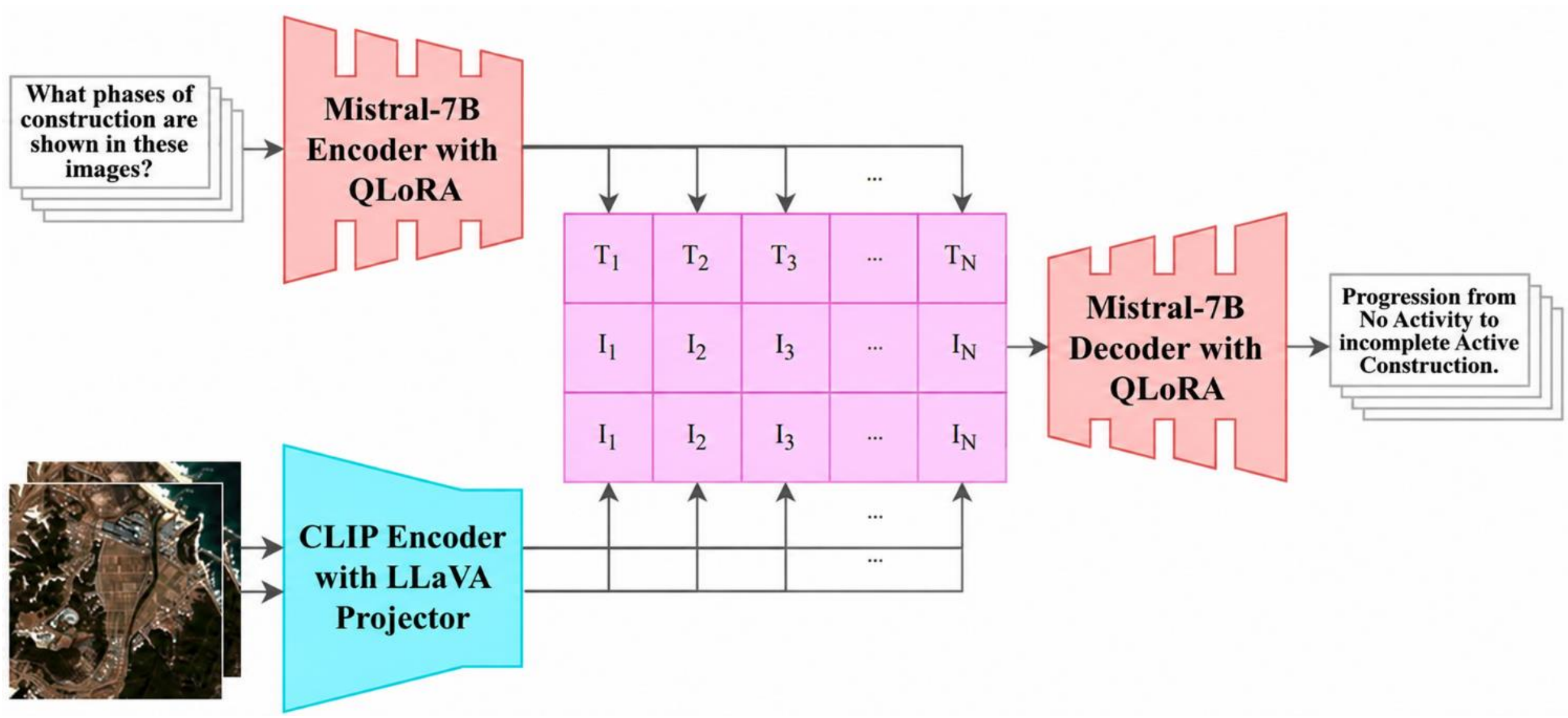

Figure 1. Proposed multi-image MLLM architecture for bi-temporal change-captioning and sensemaking of observations.

This study uses Sentinel-2 imagery and IARPA SMART-HC data labels as the primary sources to develop a related visual question answering (VQA) dataset. Sentinel-2, operated by the European Space Agency as part of the European Union Copernicus program [2], provides openly accessible, multispectral Earth-observation imagery with repeated global coverage, making it a practical foundation for reproducible MLLM research given the sparse geospatial and temporal observations. The IARPA SMART program, introduced by Goldberg et al. [3], was designed to advance automated detection, characterization, and monitoring of anthropogenic activity at a global scale using heterogeneous satellite imagery. A central motivation of SMART is that activity understanding requires spatial and temporal reasoning beyond conventional object detection or bi-temporal change detection.

Building on these foundations, we formulate SMART-HC as a temporally extended automatic target recognition (ATR) and VQA problem. In this setting, the target is not a vehicle chip or a static object crop, but a fixed geospatial construction site whose visibility changes across sparse satellite observations. We extract Sentinel-2 site chips from SMART-HC annotations, convert construction-type labels, temporal phase labels, location metadata, and observation relationships into natural-language VQA language samples, and propose a multi-image MLLM framework for predicting construction-site attributes, temporal state, and activity progression.

### 1.1 Sensemaking Definition

We define sensemaking as the process of transforming diverse, incomplete, and heterogeneous observations into actionable understanding. In remote sensing, this means moving beyond pixel-level detection, object classification, or isolated change detection toward interpretation of activity, progression, and significance. This distinction is important because modern satellite, radar, video, tracking, and detection systems can generate large volumes of sensor data. Still, the volume of collected data does not, in itself, produce meaningful intelligence. For SMART-HC analysis [1, 3], geospatial-temporal sensemaking requires integrating site imagery, metadata, construction-type labels, and temporal phase observations to infer the current state and likely progression of construction activity. The desired output is therefore not only a class label, but a concise analytic description that supports activity-based intelligence, such as determining whether a site is being prepared, actively constructed, completed, or likely associated with a specific infrastructure function. In this way, the task is closely related to temporal reasoning and intent recognition: the model must interpret multiple observations as evidence of an unfolding activity rather than as independent images.

### 1.2 Problem Formulation

Prior research has established strong benchmarks for static geospatial perception, multi-temporal change detection, remote-sensing visual question answering (VQA), and synthetic aperture radar (SAR) automatic target recognition (ATR). However, comparatively limited work has examined language-generative, multi-image multimodal large language models (MLLMs) for sparse geospatial-temporal activity sensemaking, particularly for activity sequences [1, 3]. This paper formulates geospatial-temporal sensemaking as the problem of interpreting sparse remote-sensing observations to infer the state, progression, and likely purpose of human-caused activity on the Earth's surface. Satellite imagery, radar, and other remote-sensing modalities provide repeated measurements of the same locations, but these observations are often naturally irregular, incomplete, and separated by substantial temporal gaps.

The central capability gap is therefore not simply detecting that change has occurred. Instead, the goal is to understand what activity is unfolding, what state the activity currently occupies, how the activity has changed across observations, and what future state may be expected. In the SMART-HC setting, this means treating a construction site as a fixed geospatial target whose visual characteristics evolve from land clearing and site preparation to active construction and eventual completion. The task is to predict site attributes, activity phase, temporal progression, and possible intent from one or more dated observations. Example language outputs may include "beginning construction of a large site," "active construction of an industrial structure," or "this completed building is likely a data center." This formulation reframes construction monitoring as a temporally extended ATR and VQA problem in which the model must reason across sparse image sequences rather than classify isolated images.

### 1.3 Contributions and Scope

This paper presents an ongoing effort to transform the IARPA Space-Based Machine Automated Recognition Technique Heavy Construction (SMART-HC) annotations into a language-supervised geospatial-temporal dataset for multimodal large language model (MLLM) research. The first contribution is the derivation of SMART-HC-VQA, a Sentinel-2-based visual question answering (VQA) dataset in which site-level construction observations, construction-type labels, temporal phase labels, location metadata, and image chips are converted into natural-language question-answer triplets. The second contribution is a reproducible implementation workflow for acquiring Sentinel-2 imagery [2] from public Spatio-Temporal Asset Catalog (STAC) sources, resolving heterogeneous SMART-HC imagery identifiers, chipping large satellite tiles into site-centered image samples, and preserving traceability between original annotations, downloaded imagery, and generated chips. The third contribution is a dataset analysis of the accessible Sentinel-2 subset, including construction-site chip sizes, observation counts per site, temporal coverage, construction-type imbalance, and temporal phase-label distributions. The fourth contribution is a data augmentation strategy that expands limited labeled observations into single-image VQA samples and large-scale two-image temporal-comparison examples using a novel Image-Pairwise Combinatorial Augmentation algorithm. Finally, this paper proposes a specialized multi-image MLLM architecture based on prior LLaVA-NeXT Mistral-7B automatic target recognition (ATR) work, adapted to reason over multiple dated construction-site observations and generate natural-language predictions of target attributes, temporal state, and activity progression. Together, these contributions provide a practical path from SMART-HC geospatial annotations [1, 3] to MLLM-based temporal sensemaking, while humbly avoiding claims of operational performance.

## 2. RELATED WORK

Recent research has begun to address temporal remote-sensing understanding through several complementary directions, including spatio-temporal electro-optical (EO) and synthetic aperture radar (SAR) foundation models, multi-temporal Earth-observation dialogue systems, long-term urban-dynamics benchmarks, and bi-temporal change-captioning multimodal large language models (MLLMs). These methods demonstrate growing interest in moving beyond static scene interpretation toward models that can represent change, temporal order, and evolving geospatial context. However, most existing work emphasizes representation learning, dense urban change analysis, paired-image change description, or general remote-sensing dialogue. Comparatively less work has focused on sparse, irregular, site-level activity progression and phase reasoning, where a model must interpret multiple temporally separated observations of the same geospatial target and express that progression in natural language.

### 2.1 Geospatial-Temporal MLLM

Recent remote-sensing multimodal large language models (MLLMs) have begun to address temporal understanding beyond static scene captioning and single-image visual question answering. DynamicVL [4] is one of the most directly relevant efforts, introducing a vision-language benchmark and model framework for long-term urban dynamics from multi-temporal remote-sensing imagery. Its tasks include change analysis, regional captioning, dense temporal captioning, and urban-evolution reasoning, making it a close comparator for language-mediated interpretation of geospatial change over time. BTCChat [5] addresses a narrower but highly relevant form of temporal reasoning: bi-temporal remote-sensing change captioning. By comparing paired images and explicitly modeling changed regions, BTCChat demonstrates how

MLLMs can generate natural-language descriptions of observed temporal differences, although its formulation is closer to a two-image change description than a sparse, multi-step activity progression. EarthDial [6] provides a broader Earth-observation dialogue framework that supports multi-sensor, bi-temporal, and multi-temporal interactions, positioning temporal change understanding as part of a general conversational interface for satellite imagery. Collectively, these methods show that remote-sensing MLLMs are moving toward temporal interpretation; however, they primarily emphasize long-term urban dynamics, paired-image change captioning, or broad Earth-observation dialogue. In contrast, SMART-HC sensemaking requires sparse, irregular, site-level temporal reasoning over activity phase progression [1, 3], which remains less explored.

### 2.2 Temporal Earth Foundation Models

Temporally aware Earth-observation foundation models provide a complementary line of related work to MLLM-based remote-sensing dialogue systems. SkySense [7] is particularly relevant because it treats remote sensing as a multimodal spatio-temporal representation-learning problem, using large-scale optical and synthetic aperture radar (SAR) image sequences to learn transferable Earth-observation features. Rather than focusing on language generation, SkySense emphasizes robust spatio-temporal encoding across sensors, geographies, and downstream tasks, demonstrating the value of temporal context for general-purpose remote-sensing interpretation. OlmoEarth [8] similarly advances this foundation-model direction by framing Earth-observation data as inherently spatial, multimodal, and sequential. Its stable latent image modeling approach is designed to learn reusable representations from large volumes of multimodal Earth-observation data, including temporally structured observations. These methods show that temporal structure is becoming increasingly central to the foundation of remote-sensing models [7, 8]. However, they primarily address representation learning and downstream adaptation rather than language-mediated explanation of activity. In contrast, our research uses a generative MLLM to reason over sparse site-level construction observations, where temporal phase progression may be expressed directly in natural language.

### 2.3 Remote Sensing VLM and MLLM

A growing body of work adapts vision-language models (VLMs) and multimodal large language models (MLLMs) to remote-sensing image understanding. In November 2023, GeoChat [9] first introduced grounded remote-sensing dialogue with image-level and region-level reasoning, visual grounding, captioning, visual question answering (VQA), scene classification, and language-guided object localization. RS-LLaVA [10] similarly adapts the LLaVA architecture to remote-sensing imagery for joint captioning and VQA using low-rank adaptation and remote-sensing-specific instruction data. EarthGPT [11] expands this direction into a universal multi-sensor remote-sensing MLLM by unifying optical, synthetic aperture radar (SAR), and infrared interpretation tasks through large-scale instruction tuning. At the same time, LHRS-Bot [12] incorporates volunteered geographic information and remote-sensing-specific alignment strategies to improve geospatial understanding. In parallel, DOFA [13] represents a complementary foundation-model approach, emphasizing sensor-flexible Earth-observation representation learning across heterogeneous modalities rather than language generation. More recent systems extend remote-sensing MLLMs toward knowledge grounding and scale: RS-RAG [14] retrieves external geospatial and world knowledge to improve captioning, classification, and VQA, while GeoLLaVA-8K [15] addresses ultra-high-resolution remote-sensing interpretation through token-pruning and token-selection mechanisms. Collectively, these methods establish the feasibility of adapting VLMs and MLLMs to remote-sensing imagery [9–15]. However, most focus on static image understanding, grounding, retrieval, large-scene perception, or multi-sensor interpretation rather than sparse temporal activity progression over fixed geospatial sites.

### 2.4 Advances in Automatic Target Recognition

Recent work has begun adapting vision-language models (VLMs) and multimodal large language models (MLLMs) from general remote-sensing interpretation to automatic target recognition (ATR), including fine-grained target classification, open-vocabulary detection, and interpretable decision support. For electro-optical maritime ATR, Lan et al. [16] introduce an efficient prompt-tuning strategy for large VLMs using hierarchical, multi-granularity prompts and remote-sensing ship priors to improve few-shot and unseen-category fine-grained ship classification. IFShip [17] extends this ship-recognition direction by constructing a domain-knowledge-enhanced chain-of-thought prompt generation mechanism and an instruction-following dataset for fine-grained ship classification, enabling both class prediction and natural-language reasoning. CARP [18] addresses robustness under degraded sensing conditions by developing cloud-adaptive prompting for ship classification under cloud occlusion. For broader remote-sensing object detection, VLPRSDet [19] incorporates vision-language representations into an object detector for language-conditioned recognition. At the same time, LLaMA-Unidetector [20] decouples class-agnostic localization from MLLM-based category recognition to support open-vocabulary detection. VectorLLM [21] extends this trend from object detection to structured geospatial target representation by generating vectorized building contours from remote sensing imagery, making it especially relevant to understanding constructed infrastructure. Together, these electro-optical methods show that VLMs and MLLMs can support target recognition, target localization, and structured target description in remote-sensing imagery [16–21]. However, most remain focused on single-image recognition or detection rather than temporal activity progression.

### 2.4.1 SAR ATR

Synthetic aperture radar (SAR) and SAR/electro-optical hybrid ATR research has also begun to incorporate language-supervised and multimodal reasoning. Wang et al. [22] propose a zero-shot SAR target recognition framework that uses a visual-language model to extract semantic information from optical remote-sensing imagery, a generative diffusion model to support three-dimensional target modeling and SAR simulation, and domain adaptation to reduce the gap between simulated and measured SAR imagery. SARVLM [23], including its SARCLIP contrastive component, constructs large SAR image-text corpora and trains SAR-specific vision-language models using contrastive language-image alignment, supporting semantic retrieval and zero-shot SAR target recognition. Popeye [24] is relevant as a multi-source ship detection framework that uses a unified visual-language model to support ship detection across heterogeneous remote-sensing imagery. Collectively, these methods indicate that VLMs and, to a lesser extent, MLLMs are beginning to support ATR through prompt tuning, instruction tuning, EO-to-SAR semantic transfer, SAR-specific image-text pretraining, open-vocabulary recognition, and multi-source target detection [20, 22–24]. However, they primarily address single-observation target recognition, retrieval, or localization, whereas the present work focuses on multi-observation temporal sensemaking.

### 2.4.2 MLLM-Based SAR ATR

Our prior work reframed automatic target recognition (ATR) as a language-mediated target-understanding problem rather than only a closed-set image-classification task. In our initial study [26] utilizing the Moving and Stationary Target Acquisition and Recognition (MSTAR) dataset [25], synthetic aperture radar (SAR) vehicle image chips were treated as ATR targets whose class labels and target characteristics could be converted into natural-language captions and visual question answering (VQA)-style supervision. A LLaVA-NeXT Mistral-7B vision-language model [27] was adapted using quantization and Low-Rank Adaptation (LoRA)-based parameter-efficient fine-tuning (PEFT) to predict vehicle type and descriptive target attributes from limited SAR imagery. This demonstrated that multimodal large language models (MLLMs) can learn fine-grained SAR target semantics from sparse labeled data, while also exposing practical challenges involving tokenization, overfitting, hallucinated generations, and the need for richer VQA augmentation.

The follow-on SAR-RAG work [28] extended this formulation by embedding MSTAR image chips and metadata into a vector database, enabling retrieval of semantically similar SAR exemplars as grounded context for VQA, target recognition, explanation generation, and quantitative physical-property estimation. This enabled evaluation beyond categorical accuracy, including descriptive target qualities and regression-style estimates of vehicle weight, dimensions, and weapon-system armament. Together, these studies show progression from SAR target classification to MLLM-based target characterization, in which detected objects are represented as language-supervised entities with predictable qualitative and quantitative attributes. The present SMART-HC work transfers this VQA-oriented MLLM paradigm from vehicle-centered SAR ATR to geospatial construction-site targets, where natural-language labels describe both target attributes and temporal activity state.

## 2.5 Related Datasets and Tasks

Recent SAR vision-language datasets indicate that remote-sensing image understanding is moving from closed-set classification toward instruction-following, captioning, detection, and visual question answering (VQA). SARLANG-1M [29] introduces a large-scale benchmark for SAR image-text learning, with more than one million SAR image-text pairs, fine-grained semantic descriptions, land-cover categories, object labels, and multitask question-answer pairs. SARChat-Bench-2M [30] similarly develops a large multimodal dialogue benchmark for SAR interpretation, including image-text pairs, target annotations, visual understanding tasks, and object detection tasks. These datasets are relevant because they show how remote-sensing annotations can be converted into language-supervised examples suitable for training VLMs and MLLMs. However, they are primarily designed for SAR image interpretation rather than sparse geospatial-temporal construction sensemaking. In the present work, we apply a similar language-supervision principle to the SMART-HC dataset, converting construction-site metadata, temporal phase labels, and Sentinel-2 observations into VQA examples for site-level activity reasoning.

Table 1. IARPA SMART Heavy Construction: Data Breakdown

| | Total | Sentinel-2 | Landsat-8 | WorldView | No Source |
|---|---|---|---|---|---|
| **Construction Sites** | 32,433 | 837 | 649 | 1,150 | 31,278 |
| **Target Observations** | 119,453 | 24,775 | 12,485 | 19,637 | 62,556 |
| **Source Image Tiles** | 13,254 | 5,818 | 2,328 | 5,108 | 0 |

## 3. DATA COLLECTION

The preceding related work shows that remote-sensing datasets can be transformed into language-supervised resources for captioning, dialogue, and visual question answering (VQA). In this work, we apply that principle to the IARPA SMART Heavy Construction (SMART-HC) dataset [1, 3], which provides structured geospatial-temporal annotations for construction activity rather than conventional single-image labels. Our goal is not to create new construction labels or simulate additional imagery, but to convert existing SMART-HC annotations into a reproducible MLLM-ready dataset. This section describes the source annotations, available imagery, Sentinel-2 access workflow, site-centered image-chip generation, and VQA conversion process used to derive SMART-HC-VQA.

### 3.1 IARPA SMART Heavy Construction Dataset

The IARPA SMART Heavy Construction (SMART-HC) dataset [1, 3] was created to support research on global-scale detection, classification, and monitoring of large-scale anthropogenic activity from satellite imagery. The public SMART-HC repository [31] describes the dataset as a collection of spatio-temporal annotations for heavy construction activity, intended for algorithm development and evaluation in broad-area search, activity classification, and activity prediction. Unlike datasets that localize only discrete objects, SMART-HC defines heavy construction as the construction of large-scale buildings and associated infrastructure. Site boundaries are intended to include the full spatial extent of construction-related activity, including land preparation, supporting infrastructure, access roads, parking lots, staging areas, and other associated facilities, rather than only final building footprints. This makes SMART-HC especially useful for geospatial-temporal sensemaking because its annotations describe not only where a construction site exists, but also how its spatial boundary and activity state evolve over time.

SMART-HC organizes annotations into regions, sites, and observations. A region is a large geographic area that defines the spatial and temporal bounds for annotation and processing, and may contain any number of construction sites. A site is a smaller geospatial area, larger than 8,000 $m^2$ for the Heavy Construction dataset, that represents the fundamental unit of activity. An observation is a dated image or set of images capturing the state of a site at a discrete time, with one day used as the finest temporal unit. These concepts are encoded in GeoJSON region and site models: region models define searchable spatial-temporal extents, while site models represent activity over time using dated observations, polygons, and phase labels. Because construction activity can expand, contract, or exhibit multiple simultaneous states, site boundaries may vary across observations, and sub-site polygons may be used when different portions of the same site are in different construction phases.

This annotation structure provides the information needed to convert SMART-HC into a language-supervised dataset. Region and site models provide geographic context; polygons define construction-site target boundaries; observation dates establish temporal order; construction-type labels describe site function; and temporal phase labels describe activity state. These fields naturally support VQA and instruction-following examples, including single-image questions about construction type or phase, paired-image questions about change and progression, and multi-image prompts asking a model to summarize a site's development from No Activity through Site Preparation, Active Construction, and Post Construction. In this work, SMART-HC is therefore treated not only as an activity-detection benchmark [1, 3] but as a structured source of supervision for training MLLMs to reason about fixed construction targets whose attributes and activity states evolve across sparse satellite observations.

Table 2. IARPA SMART Heavy Construction: Building Type Labels by Observation

| | Total | Sentinel-2 | Landsat-8 | WorldView | No Source |
|---|---|---|---|---|---|
| **Totals** | 121,472 | 25,628 | 12,970 | 20,285 | 62,589 |
| **Commercial** | 10,081 | 4,826 | 1,803 | 3,179 | 273 |
| **Industrial** | 7,886 | 2,736 | 1,650 | 3,347 | 153 |
| **Medium Residential** | 5,378 | 2,695 | 985 | 1,595 | 103 |
| **Heavy Residential** | 3,983 | 1,949 | 946 | 1,035 | 53 |
| **Other Type** | 2,015 | 990 | 440 | 549 | 36 |

### 3.1.1 Construction Type and Phase Annotations

SMART-HC provides two label families that are especially useful for geospatial-temporal visual question answering: construction-type labels and temporal phase labels. Construction-type labels describe the apparent functional category of a positive heavy-construction site, including Commercial, Industrial, Medium Residential, Heavy Residential, Other, Unknown, and Unlabeled records. Commercial examples may include malls, hospitals, stadiums, offices, hotels, storage facilities, and other public-facing facilities. Industrial examples may include factories, power plants, warehouses, distribution centers, manufacturing sites, and shipping or logistics infrastructure. The residential categories distinguish medium-density low-rise housing from larger high-rise residential construction. At the same time, the Other class captures heavy-construction sites that do not fit cleanly into the primary categories. Although construction-type prediction was not the central SMART-HC evaluation task [1], these labels are valuable for MLLM-based sensemaking because they convert construction observations into semantically meaningful target attributes. Site type helps contextualize the emerging infrastructure, the reasons the activity may be occurring, and the expected final site function. The construction-type labels are imbalanced, with many records marked Unlabeled. However, the labeled subset still supports direct VQA prompts such as "What type of construction site is shown?" or "Is this site commercial, industrial, or residential?", even when the ground-truth answer is Unknown. SMART-HC also defines negative, excluded, and ignored categories, including light residential construction, standalone roads or parking lots, sports fields, golf courses, solar fields, natural-disaster change, demolition without continued construction, and activities smaller than 8,000 $m^2$. These categories provide useful contrastive supervision because not every visible change, cleared lot, or site-like polygon should be interpreted as positive heavy construction. These negative, excluded, and ignored labels might be usable given additional development.

The temporal phase labels are the most important SMART-HC annotations for reasoning because they convert each construction site from a static target into an evolving activity sequence. SMART-HC defines five temporal labels: No Activity, representing the pre-construction baseline; Site Preparation, representing land clearing, grading, ground shaping, and related preparation; Active Construction, representing visible construction of buildings, foundations, intermediate structures, or supporting infrastructure; Post Construction, representing apparent completion of construction activity within the site or sub-site boundary; and Unknown, representing observations where the phase cannot be reliably determined from available imagery or supporting information. Because these labels are assigned to dated site or sub-site polygons, each construction location can be represented as a sequence of spatial boundaries and activity states over time. This enables VQA tasks that ask whether activity has begun, whether construction is progressing, whether a site has transitioned between phases, and whether it has reached an end state. The aggregated SMART-HC records contain 60,437 temporal-phase labels, dominated by 30,676 Active Construction observations; the Sentinel-2 subset contains 26,587 temporal labels, including 13,820 Active Construction observations. Although the phase labels are imbalanced, they provide direct supervision for phase classification, image-pair comparison, activity-progression description, and future-state reasoning. Together, the construction-type, negative/excluded, and temporal phase annotations extend SMART-HC beyond detection into richer target understanding, making it a structured source of language supervision for geospatial-temporal VQA.

Table 3. IARPA SMART Heavy Construction: Temporal Phase Labels by Observation

| | Total | Sentinel-2 | Landsat-8 | WorldView | No Source |
|---|---|---|---|---|---|
| **Totals** | 60,437 | 26,587 | 13,295 | 20,555 | 0 |
| **No Activity** | 7,807 | 2,497 | 1,626 | 3,684 | 0 |
| **Site Preparation** | 7,843 | 3,608 | 1,647 | 2,588 | 0 |
| **Active Construction** | 30,676 | 13,820 | 6,577 | 10,279 | 0 |
| **Post Construction** | 7,628 | 2,643 | 1,182 | 3,803 | 0 |

### 3.1.2 Satellite Imagery Sources and Access

The SMART-HC annotations are associated with a multi-source satellite-imaging ecosystem rather than a single-image product. The primary annotated sources include Landsat-8, Sentinel-2, and Maxar WorldView imagery, while Planet Labs imagery was used as a supplemental program source but is not distributed with the released annotation dataset. This multi-sensor design reflects the SMART-HC objective of combining complementary spatial, spectral, and temporal characteristics: Landsat-8 provides broad multispectral coverage at coarser spatial resolution, Sentinel-2 provides openly accessible multispectral imagery with moderate spatial resolution and frequent revisit [2], WorldView provides very-high-resolution commercial imagery, and Planet imagery offers high temporal cadence but limited public availability. In principle, this diversity supports robust geospatial-temporal activity analysis across heterogeneous sensing conditions. In practice, however, the usable imagery subset is constrained by licensing, catalog access, historical metadata consistency, and reproducibility.

The aggregated SMART-HC annotation inventory contains 119,453 target observations, 13,254 source image tiles, and 32,433 construction sites. The "No Source" category contains 62,556 target observations and 31,278 construction sites, but no associated source image tiles, making those records unavailable as direct visual inputs for MLLM training. Public access is also uneven among imagery sources. Sentinel-2 Level-2A is the most practical open source for reproducible dataset construction, while some Landsat-8 and Sentinel-2 Level-1C products may require requester-pays cloud access. WorldView and Planet imagery generally require separate commercial licensing, limiting their immediate use in an openly reproducible workflow. In addition, some early SMART-HC image references predate the current Spatio-Temporal Asset Catalog (STAC) workflows and may be difficult to resolve in modern public catalogs. Consequently, although SMART-HC was designed for multi-source geospatial-temporal analysis, this work prioritizes Sentinel-2 imagery because it offers the best balance of public accessibility, temporal coverage, reproducibility, and scale for constructing a SMART-HC-derived VQA dataset.

## 3.2 Derived Dataset: SMART-HC-VQA

The SMART-HC dataset provides an opportunity to extend prior multimodal large language model (MLLM)-based automatic target recognition (ATR) work from vehicle-centered recognition to geospatial-temporal understanding of stationary infrastructure. In this setting, a construction site is treated as an ATR target because it is a localized object or activity of interest whose identity, attributes, state, and evolution must be inferred from remote-sensing observations. Rather than recognizing a single vehicle chip or static object crop, the task is to characterize a fixed geospatial site across time. We therefore derive SMART-HC-VQA, a Sentinel-2-based visual question answering (VQA) dataset constructed from SMART-HC annotations. This dataset converts construction-type labels, temporal phase labels, geographic metadata, site observations, and observation relationships into natural-language question-answer pairs. The resulting formulation allows an MLLM to answer questions about target attributes, such as whether a site is commercial, industrial, or residential, and temporal activity states, such as No Activity, Site Preparation, Active Construction, or Post Construction. In doing so, SMART-HC-VQA reframes heavy-construction monitoring as a temporally extended ATR and VQA problem in which the target is a construction site whose observable state changes across sparse satellite observations.

Table 4. SMART-HC-VQA Sentinel-2 Data Breakdown

| Sentinel-2 Subset | Dataset Projected | Available Data | Percent Acquired |
|---|---|---|---|
| Construction Sites | 837 | 730 | 87.2% |
| Target Observations | 24,775 | 21,837 | 88.1% |
| Source Image Tiles | 5,818 | 4,839 | 83.2% |
| **Construction Type Labels** | | | |
| Commercial | 4,826 | 4,202 | 87.1% |
| Industrial | 2,736 | 2,868 | 104.8%* |
| Medium Residential | 2,695 | 2,240 | 83.1% |
| Heavy Residential | 1,949 | 1,936 | 99.3%* |
| Other Type | 990 | 1028 | 103.8%* |
| **Temporal Phase Labels** | | | |
| No Activity | 2,497 | 2,017 | 80.8% |
| Site Preparation | 3,608 | 2,595 | 71.9% |
| Active Construction | 13,820 | 11,311 | 81.8% |
| Post Construction | 2,643 | 1,444 | 54.6% |

*Some Observations include several building types labeled at different subsites.

### 3.2.1 Sentinel-2 Imagery Download

Sentinel-2 imagery was acquired from the Copernicus Data Space Ecosystem, part of the European Union Copernicus program operated by the European Space Agency (ESA) [2]. The Sentinel-2 constellation provides multispectral Earth-observation imagery with 13 spectral bands, up to 10 m spatial resolution, a 290 km swath, and approximately five-day revisit using the two-satellite constellation. When available, we use Level-2A products, which provide atmospherically corrected surface-reflectance imagery suitable for analysis.

Although Sentinel-2 data is publicly available, retrieving the imagery associated with SMART-HC annotations required a custom download workflow. The SMART-HC records reference Sentinel-2 imagery using several inconsistent formats, including EarthSearch identifiers, ESA product names, timestamp-only entries, compact tile/date strings, and SMART-specific labels that combine country, region, site, tile, date, processing level, and asset information. The downloader therefore normalizes each reference into searchable fields such as satellite platform, tile, date, processing level, site geometry, and preferred image asset. It then searches the EarthSearch Spatio-Temporal Asset Catalog (STAC) using increasingly flexible strategies, beginning with exact item lookup and then expanding to tile/date searches, fallback metadata, geometry-based searches, wider time windows, and Level-2A with Level-1C fallback when needed.

Additional engineering was required to make large-scale retrieval reliable and reproducible. Some site geometries are small and may not reliably intersect the full Sentinel-2 scene footprint, some catalog records use different names for visual image assets, and some older references are no longer available in current public catalogs. The workflow therefore ranks candidate scenes using metadata agreement and temporal proximity, retries failed downloads, avoids duplicate downloads, assigns deterministic filenames, and records traceability from each SMART-HC record to the resolved Sentinel-2 image and saved file. Using this workflow, hundreds of thousands of search permutations were issued to the EarthSearch STAC catalog, resulting in more than 1 TB of Sentinel-2 imagery for downstream chipping and quality assessment.

### 3.2.2 Construction Site Target Chipping

After Sentinel-2 source imagery was downloaded, each large satellite tile was converted into smaller site-centered image chips for visual question answering. This step was necessary because a single Sentinel-2 tile covers a large geographic area and may contain many SMART-HC construction sites observed at different times. The chipping pipeline uses each site's polygon geometry to locate the construction target, then extracts a square image centered on the site boundary. Rather than tightly masking the polygon, the chip preserves surrounding context, including roads, staging areas, nearby buildings, bare ground, and supporting infrastructure, which may be useful for interpreting construction type and phase.

Several practical geospatial issues had to be addressed during chipping. SMART-HC site geometries are provided as polygon or multi-polygon coordinates, sometimes stored in tabular formats that require parsing and validation. Because latitude and longitude coordinates do not correspond to equal ground distances, the pipeline converts each site geometry into a local projected coordinate system before defining the chip extent. The source raster is then read efficiently, resampled to a consistent 10 m ground sampling distance, and written as a north-up image chip. If the requested chip exceeds the available Sentinel-2 tile, the missing region is padded to maintain the output size and spatial extent.

The generated filenames encode key metadata, including site identifier, acquisition timestamp, construction-type label, and temporal phase label. The pipeline also records diagnostic information, including chip center, spatial bounds, pixel dimensions, ground sample distance, site span, and whether padding was required. These design choices make the chipping process reproducible and auditable while preserving the spatial context needed for downstream MLLM training. The resulting chips serve as visual inputs to SMART-HC-VQA, where each image can be paired with questions about construction type, temporal phase, location, or change over time.

### 3.2.3 Site Data Analysis

The resulting SMART-HC-VQA Sentinel-2 image-chip dataset is uneven at the construction-site level, which has important implications for both temporal learning and MLLM training efficiency. Most sites produce relatively small chips, with image dimensions concentrated around 50 × 50 pixels. Only a small number of sites exceed 100 × 100 pixels, a few exceed 200 × 200 pixels, and rare outliers exceed 400 × 400 pixels. This long-tailed size distribution creates two competing challenges: very small chips may lack sufficient visual context for reliable interpretation, while large chips require many more visual tokens and may produce GPU memory spikes during training. Observation counts are similarly imbalanced. Many construction sites have only one or a few usable observations, while a smaller subset contains tens of observations, and rare sites contain more than 100 or even 250 temporal image chips. Temporal coverage also varies widely, with many sites observed on only a single day and others spanning multi-year periods of up to five years. As a result, the dataset comprises sparse, irregular satellite observation histories rather than dense video-like sequences. Despite this sparsity, many sites contain multiple temporal phase labels. The most informative examples include several phases of construction, typically No Activity, Site Preparation, Active Construction, and Post Construction. Sites with only one observed phase provide limited evidence of change, while sites with three or more observed phases are especially valuable for learning temporal progression. Overall, the dataset supports geospatial-temporal sensemaking, but its distribution reveals practical challenges: most sites are small, many temporal sequences are sparse, and the most useful multi-phase construction histories are concentrated in a subset of available locations.

### 3.2.4 Generating VQA Triplets

To convert SMART-HC annotations into visual question answering (VQA) data, each labeled site observation is represented as an image-question-answer triplet. The image is the Sentinel-2 chip centered on a construction site, the question asks about a known property of that site, and the answer is the corresponding ground-truth value from the SMART-HC metadata. Construction-type labels support questions such as "What type of construction site is shown?", temporal phase labels support questions such as "What phase of construction is visible?", and location metadata supports questions about the country or region associated with the site. We define a finite set of reliable question types using metadata fields with known values, ensuring that target answers remain deterministic and directly evaluable. Each of the 21,837 Sentinel-2 construction-site chips is paired with multiple metadata-derived VQA tasks, including construction type, temporal phase, and geographic location, producing 65,511 single-image VQA samples. To improve language diversity, large language models (LLMs) can generate paraphrased versions of each question while preserving the same ground-truth answer. This increases the variety of natural-language prompts without introducing ambiguity into the output labels. In this way, SMART-HC construction types, temporal phases, locations, and related metadata are transformed into single-image VQA supervision to learn both site attributes and construction activity states.

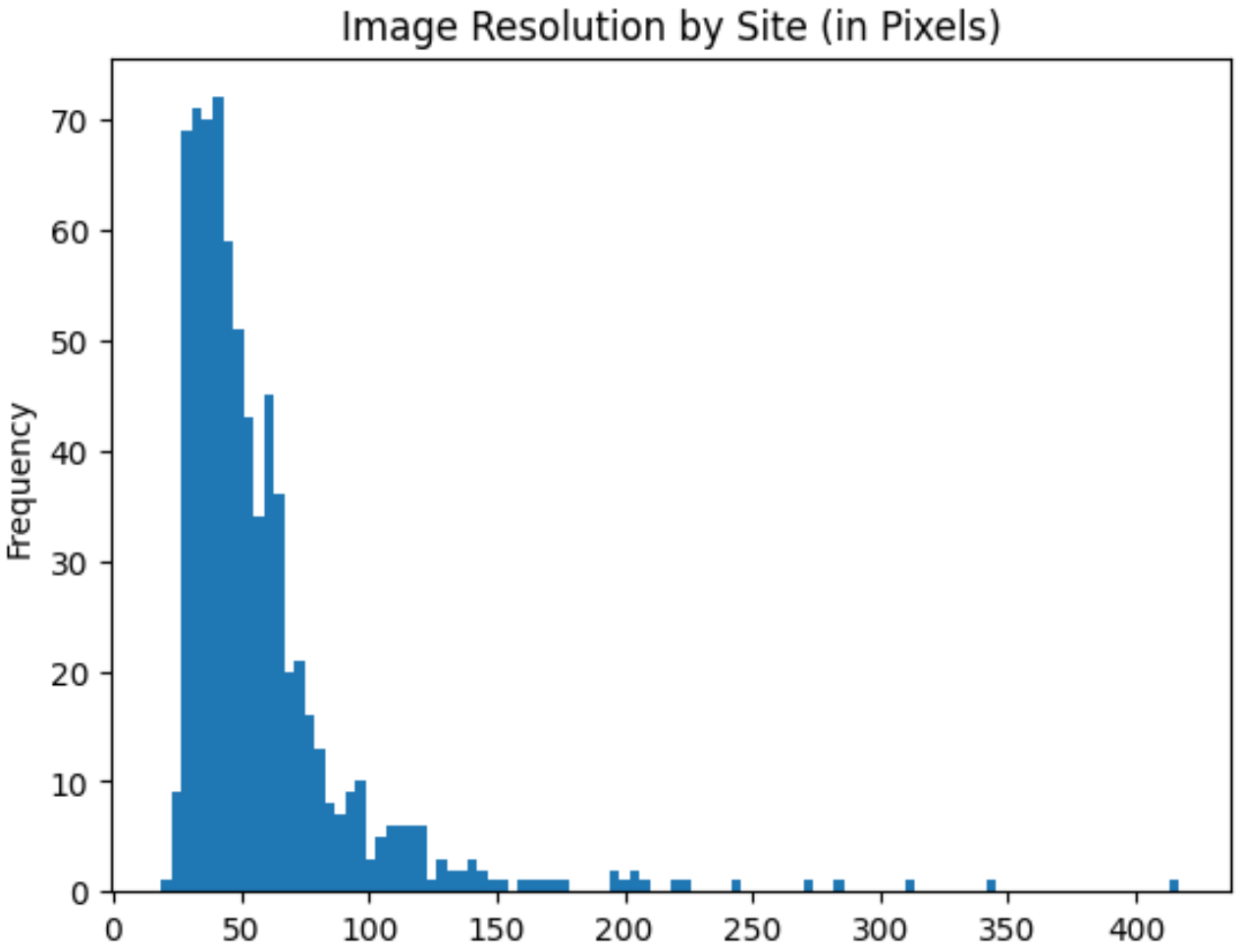

Figure 2. A histogram shows the square-chipped image resolution, in pixels, for all Sites and accompanying Observations.

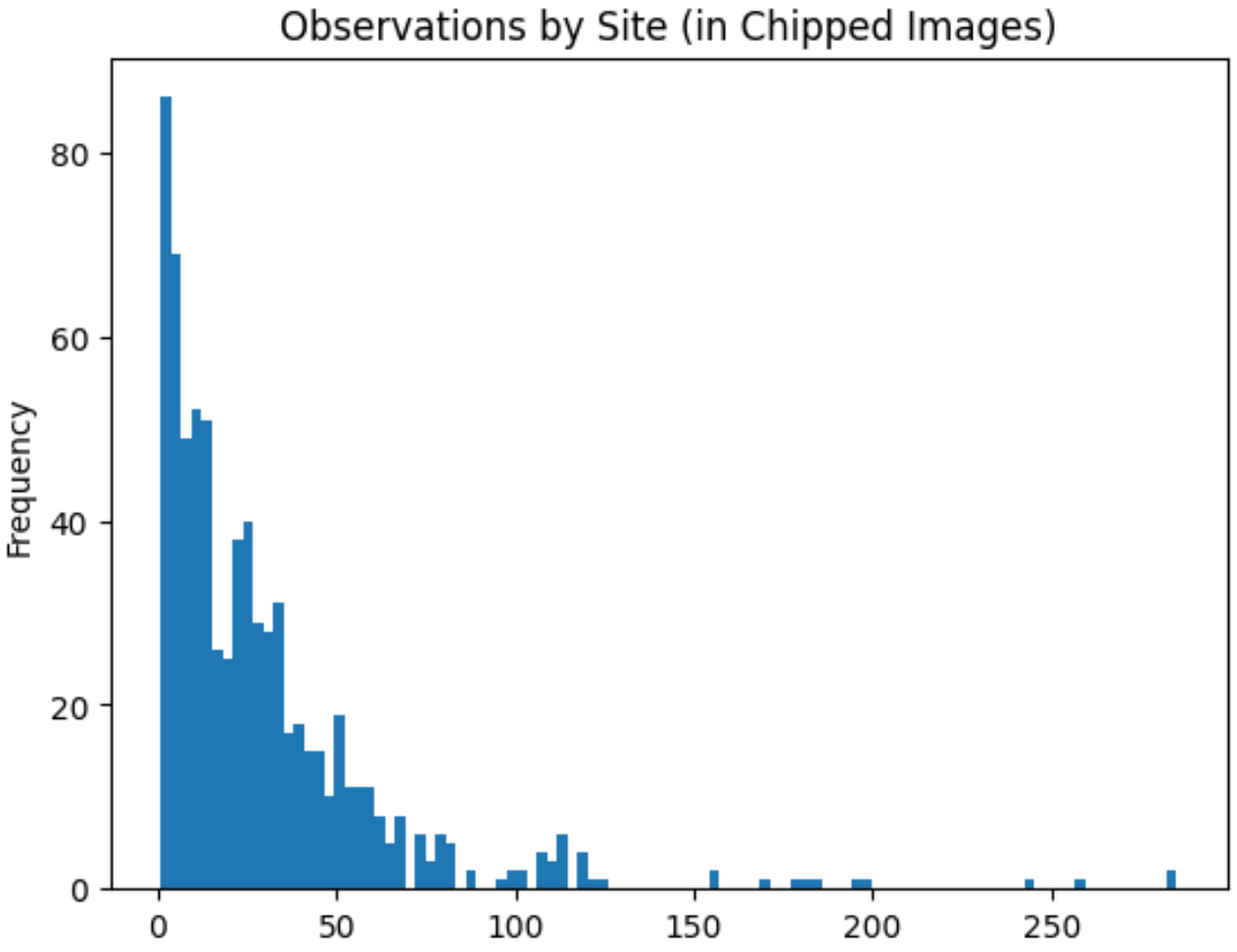

Figure 3. A histogram shows the quantity of Observations and chipped images per Site, with many having few or only one.

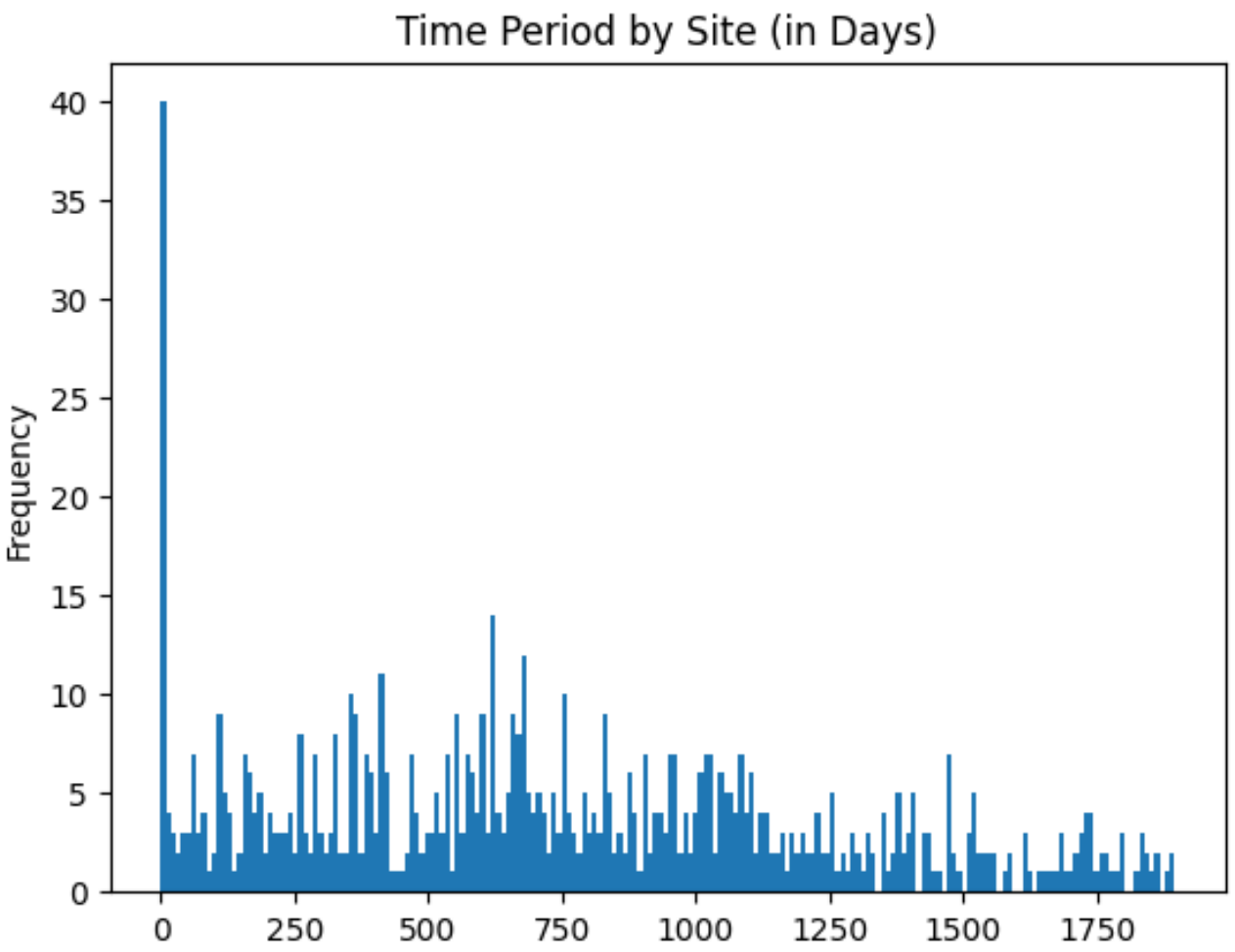

Figure 4. A histogram shows the period in days between the first and last Observation at each Site, many with only one day.

### 3.3 Image-Pairwise Combinatorial Augmentation

To extend beyond single-image VQA, we introduce Image-Pairwise Combinatorial Augmentation, which uses multiple observations from the same construction site to generate two-image temporal comparison examples. In this setting, a multi-image MLLM receives two dated chips from the same site and is asked to compare them, identify whether the site changed, determine relative temporal order, or classify the construction-state transition between the two observations. Because a site with multiple observations can be paired in many valid combinations, the number of supervised examples grows combinatorially without requiring new imagery, simulated data, or manually annotated labels. In this study, pairwise expansion increases the dataset from 65,511 single-image VQA examples to approximately 2.3 million two-image temporal comparison examples. This augmentation transforms sparse satellite observations into a large set of supervised temporal reasoning tasks, enabling future models to learn relationships such as No Activity to Site Preparation, Site Preparation to Active Construction, Active Construction to Post Construction, or no apparent phase change.

## 4. MLLM TRAINING

This work implements a multi-image multimodal large language model (MLLM) training framework for geospatial-temporal sensemaking of SMART-HC construction-site observations. The framework adapts the LLaVA-NeXT Mistral-7B automatic target recognition (ATR) pipeline from our prior synthetic aperture radar (SAR) work to the Sentinel-2 SMART-HC-VQA dataset, where each construction site is treated as a fixed geospatial target whose attributes and activity state evolve over time. The current implementation supports single-image questions about construction type, temporal phase, and location, as well as two-image questions about change, temporal ordering, and phase progression. Rather than proposing a new foundation model, this section describes how an existing MLLM architecture was modified to accept multiple dated image inputs and train on metadata-derived visual question answering (VQA) examples.

### 4.1 Baseline MLLM Architecture

The training framework builds on the LLaVA-NeXT Mistral-7B architecture used in our prior MLLM-based SAR ATR work [26–28]. This model follows a LLaVA-style design that combines a pretrained language model, a vision encoder, and a multimodal projection layer for image-conditioned language generation. A natural-language question is tokenized and processed by the Mistral-7B language model. At the same time, each input image is encoded by a Contrastive Language-Image Pre-training (CLIP) Vision Transformer (ViT)-Large Patch-14 336-pixel vision encoder. The LLaVA-NeXT projector maps the resulting visual embeddings into the language-model embedding space, where they are inserted into the prompt context. The decoder then generates an answer conditioned jointly on the question and the visual content.

This baseline is appropriate for SMART-HC-VQA because it has already been adapted to remote-sensing ATR tasks using limited labeled imagery. In our prior MSTAR work, LLaVA-NeXT Mistral-7B was trained with quantization and parameter-efficient fine-tuning rather than full model retraining. Specifically, 4-bit quantized model weights, 16-bit brain floating point computation, and Quantized Low-Rank Adaptation (QLoRA) reduced memory requirements while preserving the pretrained model's language and visual reasoning capabilities. For the present work, we retain this baseline architecture but modify the input structure and training data so that the model can process Sentinel-2 construction-site chips and answer questions about construction attributes, temporal phases, and activity progression.

### 4.2 Multi-Image Input Adaptation

The central architectural change for SMART-HC-VQA is support for multiple dated observations from the same construction site. In the baseline single-image VQA setting, the model receives a single image chip and a single text prompt. For geospatial-temporal sensemaking, however, the model must compare two or more Sentinel-2 observations of the same site and generate an answer that reflects both visual evidence and temporal context. To support this, each Sentinel-2 chip is encoded independently by the shared CLIP vision encoder and LLaVA-NeXT projector. The resulting visual-token sequences are inserted into the language-model context as separate image segments.

Each image segment is paired with explicit text markers or metadata so the language model can distinguish observations. For example, prompts may identify inputs as "Image 1" and "Image 2," include acquisition dates, or specify whether the images are in chronological order. These markers are important because the decoder does not inherently know that two image-token blocks correspond to different times unless that relationship is explicitly encoded in the prompt or learned during training. In the current implementation, temporal reasoning is therefore induced through ordered image placement, date-aware prompts, and supervised answers derived from SMART-HC phase labels, rather than through a separate temporal transformer or video encoder.

This design is well matched to sparse satellite observations. SMART-HC image sequences are not dense videos; they are irregularly spaced snapshots of the site state. The current training implementation, therefore, emphasizes two-image comparison as the first practical temporal learning task. Longer site histories may later be represented by adding additional dated image segments, selecting key observations, or retrieving prior examples from site-level memory.

### 4.3 VQA Task Formatting

Training examples are formatted as instruction-following visual question answering (VQA) samples with deterministic answers derived from SMART-HC metadata. In the single-image setting, each example contains one Sentinel-2 construction-site chip, a natural-language question, and a canonical answer for construction type, temporal phase, location, or another known site attribute. For example, a prompt may ask, "What type of construction site is shown?" or "What phase of construction is visible in this image?" with answers mapped to fixed label sets such as Commercial, Industrial, Site Preparation, Active Construction, Post Construction, or Unknown.

In the multi-image setting, each example contains two dated observations from the same construction site. The question asks the model to compare the images, identify temporal order, determine whether meaningful change occurred, or classify the phase transition between observations. Representative answer labels include no apparent phase change, No Activity to Site Preparation, Site Preparation to Active Construction, and Active Construction to Post Construction. This controlled VQA format is intentionally preferred over unrestricted narrative captioning at this stage because generated responses can be normalized to known labels and scored reproducibly. Question paraphrasing may be used to increase linguistic diversity, but the target answers remain fixed to preserve precise evaluation.

### 4.4 Parameter-Efficient Fine-Tuning

Fine-tuning a full multimodal large language model is computationally expensive because training must store model parameters, gradients, optimizer states, and intermediate activations. This cost increases in the SMART-HC-VQA setting because each additional image adds visual tokens to the language model context. As a result, multi-image examples require careful control of context length and graphics processing unit (GPU) memory usage. The current training implementation, therefore, prioritizes single-image and two-image examples before extending to longer site histories.

We follow the parameter-efficient fine-tuning approach used in our prior LLaVA-NeXT Mistral-7B automatic target recognition work. The pretrained language model and vision encoder remain largely frozen, while a small number of trainable adapter parameters are introduced through Quantized Low-Rank Adaptation (QLoRA). In the baseline configuration, model weights are loaded with 4-bit quantization, computation uses 16-bit brain floating point precision, and Low-Rank Adaptation (LoRA) modules are inserted into selected language-model attention layers. This allows the model to adapt to SMART-HC terminology, Sentinel-2 visual appearance, construction-type labels, temporal phase labels, and phase-transition patterns without retraining the full foundation model.

The prior SAR ATR implementation used a LoRA rank of 8, LoRA scaling of 16, a 5% dropout rate, and fused AdamW optimization as practical starting settings. These values are retained as initial hyperparameters for SMART-HC-VQA, but they are not assumed to be optimal. Future training experiments should evaluate adapter placement, LoRA rank, batching by image count or chip size, and visual-token reduction methods to improve stability and reduce memory spikes during multi-image fine-tuning.

### 4.5 Training and Evaluation Plan

Training is organized as a staged progression from simpler single-image tasks to more difficult temporal comparison tasks. In the first stage, the model is trained on single-image SMART-HC-VQA examples to learn the appearance of Sentinel-2 construction sites and metadata-derived labels, including construction type, temporal phase, location, and other site attributes. In the second stage, the model is trained on two-image examples from the same construction site, where it must compare dated observations, determine temporal order, identify whether meaningful change occurred, and classify phase transitions, such as No Activity to Site Preparation or Site Preparation to Active Construction. Sites with longer observation histories may later support short ordered-sequence examples, but pairwise comparison is the primary temporal training task in the current framework.

Evaluation will similarly be organized by task type. Single-image construction-type and temporal-phase questions can be evaluated as categorical classification after mapping generated text to canonical labels. Pairwise examples can be evaluated using temporal-order accuracy and transition-classification accuracy, with responses normalized to labels such as no apparent change, No Activity to Site Preparation, Site Preparation to Active Construction, Active Construction to Post Construction, or Unknown transition. Because this paper presents an early-stage framework rather than final performance results, the evaluation plan emphasizes controlled answer formats and reproducible label matching before open-ended narrative sensemaking. This conservative strategy allows SMART-HC-VQA to first function as an automatic target recognition and VQA benchmark, while later supporting richer activity summaries, quantitative site measurements, retrieval-augmented reasoning, and intent-oriented geospatial analysis.

## 5. FUTURE WORK

Future work will extend SMART-HC-VQA in three directions: richer sensing, richer supervision, and more efficient multimodal model architectures. First, the dataset can be expanded beyond Sentinel-2 optical imagery by incorporating synthetic aperture radar (SAR) observations from Sentinel-1 [32] and commercial SAR providers such as Capella Space and Umbra. SAR imagery provides all-weather, day/night sensing that may help fill temporal gaps caused by cloud cover, seasonal effects, or sparse optical collection. Combining Sentinel-2 electro-optical imagery with SAR observations would move the task from single-sensor temporal visual question answering (VQA) toward multi-sensor geospatial sensemaking, where a model must reason over complementary spatial, spectral, temporal, and radar-scattering evidence. Additional geospatial layers, including Land Use/Land Cover (LULC) products, transportation networks, zoning information, vegetation indices, and infrastructure maps, could also provide weak contextual labels for site function and surrounding activity.

Second, future work should expand the prediction targets from categorical labels to quantitative measurements and higher-level activity descriptions. The current SMART-HC-VQA formulation focuses on construction type, temporal phase, location, and phase transitions. However, SMART-HC site polygons also support quantitative questions, including construction-site area, area change over time, elapsed time between phases, and apparent expansion rate. This direction would extend the regression-style VQA approach explored in SAR-RAG [28] from vehicle measurements to fixed infrastructure targets. More broadly, construction sites should be treated as evolving geospatial targets, with observable attributes including size, shape, surrounding land use, degree of completion, and temporal state. The long-term objective is not only to classify a site as Active Construction, but to infer what activity is occurring, how it is progressing, what contextual factors may explain it, and what future state is likely.

Third, future experiments should evaluate newer multimodal foundation models as alternatives to the LLaVA-NeXT Mistral-7B baseline, especially models that improve image reasoning while remaining computationally practical. Gemma 3 [33] is relevant because it includes lightweight multimodal variants, supports long context, and introduces memory-efficient architectural changes that may benefit multi-image remote-sensing VQA. LLaMA 3.2 [34] is also relevant because its 11B vision model supports image reasoning, captioning, and visual grounding while remaining more practical than larger frontier-scale systems; its smaller text-only variants may also support metadata reasoning, question generation, or retrieval-augmented pipelines. Qwen3 [35] provides strong small dense language models that could serve as efficient language-side backbones when paired with a vision encoder and projector. Recent DeepSeek conditional-memory and V4 releases [36, 37] suggest another promising direction: separating external memory or lookup-style context from dynamic reasoning. Such memory-augmented approaches could store site histories, prior observations, geographic context, and known construction examples without forcing all information into model weights.

The most promising near-term path is therefore to combine efficient multimodal backbones with better temporal data organization, rather than simply scaling to a much larger model. Future studies should compare LLaVA-NeXT Mistral-7B against smaller or newer multimodal alternatives using identical single-image and image-pairwise VQA splits. Longer-term work should investigate retrieval-augmented site memory, Sentinel-1/Sentinel-2 SAR/EO fusion, quantitative site regression, and short ordered image sequences. These extensions support the broader objective of moving from construction-phase recognition toward temporal reasoning and intent recognition over geospatial targets, including more complex applications such as infrastructure expansion, maritime activity monitoring, and other scenarios whose meaning depends jointly on time, location, activity state, and environmental conditions.

## 6. CONCLUSIONS

This paper presents ongoing work toward geospatial-temporal sensemaking of remote-sensing activity observations using multimodal large language models. Rather than treating the IARPA SMART Heavy Construction dataset only as a benchmark for broad-area search or phase classification, we reinterpret it as a structured source of language supervision for visual question answering. In this formulation, a construction site is treated as a fixed geospatial target whose identity, attributes, and activity state evolve across sparse satellite observations. This reframes heavy-construction monitoring as a temporally extended automatic target recognition and VQA problem, where the objective is not only to classify an image, but to describe what is happening at a site, how the activity is progressing, and what the observed development may imply.

A central contribution of this work is the derivation of SMART-HC-VQA from SMART Heavy Construction annotations and Sentinel-2 imagery. We convert construction-type labels, temporal phase labels, location metadata, site observations, and image relationships into natural-language question-answer examples suitable for MLLM training. The current derived dataset includes 21,837 accessible Sentinel-2 construction-site image chips, 65,511 single-image VQA examples, and approximately 2.3 million two-image temporal comparison examples created through Image-Pairwise Combinatorial Augmentation. This expansion increases the number of supervised temporal comparison examples without requiring new simulations, synthetic imagery, or unsupported labels.

This work also highlights the practical engineering required to make public geospatial-temporal annotations usable for MLLM research. We describe a workflow for resolving heterogeneous Sentinel-2 source identifiers, retrieving imagery from public STAC catalogs, chipping large satellite tiles into site-centered images, preserving surrounding spatial context, and maintaining traceability between original SMART-HC records and generated image chips. The resulting dataset analysis shows that SMART-HC-VQA is valuable but naturally uneven: many sites are small, many contain only a few observations, temporal coverage is sparse and irregular, and construction-type and phase labels are imbalanced. These properties reflect the real challenges of satellite-based activity understanding and motivate both data augmentation and multi-image modeling.

Finally, we describe an implemented multi-image MLLM training framework based on LLaVA-NeXT Mistral-7B. The framework extends a prior SAR ATR pipeline by allowing multiple dated Sentinel-2 image chips to be inserted into the language-model context as separate visual inputs, with prompt markers and metadata used to distinguish observations. The current system is designed to train on single-image and two-image SMART-HC-VQA examples, supporting questions about construction type, temporal phase, temporal order, and phase transition. Although this paper does not present final model evaluation results, it defines a reproducible workflow for dataset construction, a controlled VQA task formulation, an augmentation strategy, and a practical training direction. Together, these contributions establish a foundation for language-guided geospatial-temporal sensemaking, with future extensions toward SAR/EO fusion, retrieval-augmented site memory, quantitative site measurement, and broader temporal reasoning over human activity on the Earth's surface.

## ACKNOWLEDGEMENTS

The authors gratefully acknowledge the IARPA SMART program, producers of the SMART Heavy Construction dataset, for making the annotations, documentation, and supporting resources available to the research community. The authors also thank the European Space Agency and the Copernicus Sentinel-2 mission for providing the publicly accessible multispectral Earth-observation imagery used in this study. This work is supported by Prime Solutions Group Inc. and the ASU Sensor, Signal, and Information Processing (SenSIP) Center. A.I. tools were used during our literature review, proofreading, and editing, including ChatGPT, Consensus, and Grammarly.

## REFERENCES

[1] C. R. Ratto, M. T. Kelbaugh, T. A. Stout, C. D. Piatko, and H. R. Goldberg, "Evaluating broad area search and classification of heavy construction activity from multi-source, multi-temporal satellite image sequences," in Geospatial Informatics XV, Proc. SPIE, vol. 13461, Art. no. 1346107, 2025, doi.org/10.1117/12.3053632

[2] European Space Agency, "Sentinel-2," Copernicus Sentinel Missions, ESA. Accessed: May 8, 2026. [Online]. Available: esa.int/Applications/Observing_the_Earth/Copernicus/Sentinel-2

[3] H. R. Goldberg, C. R. Ratto, A. Banerjee, M. T. Kelbaugh, M. Giglio, and E. F. Vermote, “Automated global-scale detection and characterization of anthropogenic activity using multi-source satellite-based remote sensing imagery,” in Geospatial Informatics XIII, Proc. SPIE, vol. 12525, Art. no. 1252502, 2023, doi.org/10.1117/12.2663071

[4] W. Xuan, J. Wang, H. Qi, Z. Chen, Z. Zheng, Y. Zhong, J. Xia, and N. Yokoya, “DynamicVL: Benchmarking multimodal large language models for dynamic city understanding,” arXiv:2505.21076, 2025, doi.org/10.48550/arXiv.2505.21076

[5] Y. Li, W. Xu, Y. Zhang, Z. Wei, and M. Peng, “BTCChat: Advancing remote sensing bi-temporal change captioning with multimodal large language model,” arXiv:2509.05895, 2025, doi.org/10.48550/arXiv.2509.05895

[6] S. Soni, A. Dudhane, H. Debary, M. Fiaz, M. A. Munir, M. S. Danish, P. Fraccaro, C. D. Watson, L. J. Klein, F. S. Khan, and S. Khan, “EarthDial: Turning multi-sensory Earth observations to interactive dialogues,” in Proc. IEEE/CVF Conf. Computer Vision and Pattern Recognition (CVPR), 2025, doi.org/10.1109/CVPR52734.2025.01334

[7] X. Guo, J. Lao, B. Dang, Y. Zhang, L. Yu, L. Ru, L. Zhong, Z. Huang, K. Wu, D. Hu, H. He, J. Wang, J. Chen, M. Yang, Y. Zhang, and Y. Li, “SkySense: A multi-modal remote sensing foundation model towards universal interpretation for Earth observation imagery,” in Proc. IEEE/CVF Conf. Computer Vision and Pattern Recognition (CVPR), 2024, pp. 27662–27673, doi.org/10.1109/CVPR52733.2024.02613

[8] H. Herzog, F. Bastani, Y. Zhang, G. Tseng, J. Redmon, H. Sablon, R. Park, J. Morrison, A. Buraczynski, K. Farley, J. Hansen, A. Howe, P. A. Johnson, M. Otterlee, T. Schmitt, H. Pitelka, S. Daspit, R. Ratner, C. Wilhelm, S. Wood, M. Jacobi, H. Kerner, E. Shelhamer, A. Farhadi, R. Krishna, and P. Beukema, “OlmoEarth: Stable latent image modeling for multimodal Earth observation,” arXiv:2511.13655, 2025, doi.org/10.48550/arXiv.2511.13655

[9] K. Kuckreja, M. S. Danish, M. Naseer, A. Das, S. H. Khan, and F. S. Khan, “GeoChat: Grounded large vision-language model for remote sensing,” in Proc. IEEE/CVF Conf. Computer Vision and Pattern Recognition (CVPR), 2024, pp. 27831–27840, doi.org/10.1109/CVPR52733.2024.02629

[10] Y. Bazi, L. Bashmal, M. M. Al Rahhal, R. Ricci, and F. Melgani, “RS-LLaVA: A large vision-language model for joint captioning and question answering in remote sensing imagery,” Remote Sensing, vol. 16, no. 9, Art. no. 1477, 2024, doi.org/10.3390/rs16091477

[11] W. Zhang, M. Cai, T. Zhang, Z. Yin, and X. Mao, “EarthGPT: A universal multi-modal large language model for multi-sensor image comprehension in remote sensing domain,” IEEE Trans. Geosci. Remote Sens., vol. 62, pp. 1–20, 2024, doi.org/10.1109/TGRS.2024.3409624

[12] D. Muhtar, Z. Li, F. Gu, X. Zhang, and P. Xiao, “LHRS-Bot: Empowering remote sensing with VGI-enhanced large multimodal language model,” in Computer Vision – ECCV 2024, Lecture Notes in Computer Science, Cham, Switzerland: Springer, 2024, pp. 440–457, doi.org/10.1007/978-3-031-72904-1_26

[13] Z. Xiong, Y. Wang, F. Zhang, A. J. Stewart, J. Hanna, D. Borth, I. Papoutsis, B. Le Saux, G. Camps-Valls, and X. X. Zhu, “Neural plasticity-inspired foundation model for observing the Earth crossing modalities,” arXiv:2403.15356, 2024, doi.org/10.48550/arXiv.2403.15356

[14] C. Wen, Y. Lin, X. Qu, N. Li, Y. Liao, H. Lin, and X. Li, “Remote sensing retrieval-augmented generation: Bridging remote sensing imagery and comprehensive knowledge with a multimodal dataset and retrieval-augmented generation model,” IEEE Geosci. Remote Sens. Mag., pp. 2–20, 2026, doi.org/10.1109/MGRS.2025.3645852

[15] F. Wang, M. Chen, Y. Li, D. Wang, H. Wang, Z. Guo, Z. Wang, B. Shan, L. Lan, Y. Wang, H. Wang, W. Yang, B. Du, and J. Zhang, “GeoLLaVA-8K: Scaling remote-sensing multimodal large language models to 8K resolution,” arXiv:2505.21375, 2025, doi.org/10.48550/arXiv.2505.21375

[16] L. Lan, F. Wang, X.-T. Zheng, Z. Wang, and X. Liu, “Efficient prompt tuning of large vision-language model for fine-grained ship classification,” IEEE Trans. Geosci. Remote Sens., vol. 63, pp. 1–10, 2025, doi.org/10.1109/TGRS.2024.3509721

[17] M. Guo, M. Wu, Y. Shen, H. Li, and C. Tao, “IFShip: Interpretable fine-grained ship classification with domain knowledge-enhanced vision-language models,” Pattern Recognition, vol. 166, Art. no. 111672, 2025, doi.org/10.1016/j.patcog.2025.111672

[18] H. Zhan, Y. Song, X. Huang, X. Tan, and T. Zhang, “CARP: Cloud-adaptive robust prompting of vision-language models for ship classification under cloud occlusion,” Frontiers in Remote Sensing, vol. 6, Art. no. 1662024, 2025, doi.org/10.3389/frsen.2025.1662024

[19] D. Liu, X. Liang, Y. Qi, Y. Xi, J. Jin, and J. Zhang, "VLPRSDet: A vision-language pretrained model for remote sensing object detection," Neurocomputing, vol. 658, Art. no. 131712, 2025, doi.org/10.1016/j.neucom.2025.131712

[20] G. Wang, J. Xie, T. Zhang, Y. Sun, H. Chen, Z. Yin, and J. Li, "LLaMA-Unidetector: A LLaMA-based universal framework for open-vocabulary object detection in remote sensing imagery," *IEEE Transactions on Geoscience and Remote Sensing*, vol. 63, pp. 1–18, 2025, doi.org/10.1109/TGRS.2025.3564332

[21] T. Zhang, S. Wei, S. Chen, W. Yu, M. Luo, and S. Ji, "VectorLLM: Human-like extraction of structured building contours via multimodal LLMs," ISPRS J. Photogramm. Remote Sens., vol. 233, pp. 55–68, 2026, doi.org/10.1016/j.isprsjprs.2026.01.025

[22] J. Wang, H. Sun, T. Tang, Y. Sun, Q. He, L. Lei, and K. Ji, "Leveraging visual language model and generative diffusion model for zero-shot SAR target recognition," Remote Sensing, vol. 16, no. 16, Art. no. 2927, 2024, doi.org/10.3390/rs16162927

[23] Q. Ma, Z. Wang, W. Liu, X. Lu, B. Deng, P. Duan, X. Kang, and S. Li, "SARVLM: A vision language foundation model for semantic understanding and target recognition in SAR imagery," arXiv:2510.22665, 2025, doi.org/10.48550/arXiv.2510.22665

[24] W. Zhang, M. Cai, T. Zhang, G. Lei, Y. Zhuang, and X. Mao, "Popeye: A unified visual-language model for multi-source ship detection from remote sensing imagery," IEEE Journal of Selected Topics in Applied Earth Observations and Remote Sensing, vol. 17, pp. 19813–19826, 2024, doi.org/10.1109/JSTARS.2024.3488034

[25] DARPA and AFRL, "Moving and Stationary Target Acquisition and Recognition (MSTAR) Public Release," Sensor Data Management System, Sep. 1995. [Online]. Available: sdms.afrl.af.mil/index.php?collection=mstar

[26] D. F. Ramirez, T. L. Overman, K. Jaskie, M. Kleine, and A. Spanias, "Towards a large language-vision question answering model for MSTAR automatic target recognition," in Automatic Target Recognition XXXV, K. Chen, R. I. Hammoud, and T. L. Overman, Eds., Proc. SPIE, vol. 13463, Art. no. 134630D, 2025, doi.org/10.1117/12.3053859

[27] B. Li, K. Zhang, H. Zhang, D. Guo, R. Zhang, F. Li, Y. Zhang, Z. Liu, and C. Li, "LLaVA-NeXT: Stronger LLMs supercharge multimodal capabilities in the wild," LLaVA Blog, May 2024. [Online]. Available: llava-vl.github.io/blog/2024-05-10-llava-next-stronger-llms/

[28] D. F. Ramirez, T. Overman, K. Jaskie, J. Marvin, and A. Spanias, "SAR-RAG: ATR visual question answering by semantic search, retrieval, and MLLM generation," arXiv:2602.04712, 2026, doi.org/10.48550/arXiv.2602.04712

[29] Y. Wei, A. Xiao, Y. Ren, Y. Zhu, H. Chen, J. Xia, and N. Yokoya, "SARLANG-1M: A benchmark for vision-language modeling in SAR image understanding," IEEE Trans. Geosci. Remote Sens., 2026, doi.org/10.1109/TGRS.2026.3652099

[30] Z. Ma, X. Xiao, S. Dong, P. Wang, H. Wang, and Q. Pan, "SARChat-Bench-2M: A multi-task vision-language benchmark for SAR image interpretation," arXiv:2502.08168, 2025, doi.org/10.48550/arXiv.2502.08168

[31] JHU/APL PubGeo, "IARPA SMART Heavy Construction Dataset," GitHub repository. Accessed: May 8, 2026. [Online]. Available: github.com/pubgeo/IARPA-SMART

[32] European Space Agency, "Sentinel-1," Copernicus Sentinel Missions, ESA. Accessed: May 8, 2026. [Online]. Available: esa.int/Applications/Observing_the_Earth/Copernicus/Sentinel-1

[33] Gemma Team, A. Kamath, J. Ferret, S. Pathak, N. Vieillard, R. Merhej, S. Perrin, T. Matejovicova, A. Ramé, M. Rivière, et al., "Gemma 3 technical report," arXiv:2503.19786, 2025, doi.org/10.48550/arXiv.2503.19786

[34] A. Dubey, A. Jauhri, A. Pandey, A. Kadian, A. Al-Dahle, A. Letman, A. Mathur, A. Schelten, A. Yang, A. Fan, et al., "The Llama 3 herd of models," arXiv:2407.21783, 2024, doi.org/10.48550/arXiv.2407.21783

[35] A. Yang, A. Li, B. Yang, B. Zhang, B. Hui, B. Zheng, B. Yu, C. Gao, C. Huang, C. Lv, C. Zheng, D. Liu, F. Zhou, F. Huang, F. Hu, H. Ge, H. Wei, H. Lin, J. Tang, J. Yang, et al., "Qwen3 technical report," arXiv:2505.09388, 2025, doi.org/10.48550/arXiv.2505.09388

[36] X. Cheng, W. Zeng, D. Dai, Q. Chen, B. Wang, Z. Xie, K. Huang, X. Yu, Z. Hao, Y. Li, H. Zhang, H. Zhang, D. Zhao, and W. Liang, "Conditional memory via scalable lookup: A new axis of sparsity for large language models," arXiv:2601.07372, 2026, doi.org/10.48550/arXiv.2601.07372

[37] DeepSeek-AI, "DeepSeek-V4: Towards highly efficient million-token context intelligence," Tech. Rep., Apr. 2026. [Online]. Available: huggingface.co/deepseek-ai/DeepSeek-V4-Pro/blob/main/DeepSeek_V4.pdf